\documentclass[12pt,oneside,english]{article}

 \usepackage{epsfig}
\usepackage[T1]{fontenc}
\usepackage[latin1]{inputenc}
\usepackage{babel}
\usepackage{setspace}
\usepackage{rotating}

\makeatletter

\begin{document}

\centerline{ \bf \large  Gravitational wave holography }

\bigskip \bigskip

\centerline{\large  D. Bar} 

\bigskip

\begin{abstract}
{ \it We represent and discuss a theory of  gravitational holography  
in which
all the involved waves; subject, reference and illuminator are  
gravitational waves (GW).   Although these waves are so weak  that no 
terrestrial experimental set-ups, even
the large LIGO, VIRGO, GEO and TAMA facilities,  were  able up to now 
to directly detect 
them  they are,
nevertheless,  known under
certain conditions (such as very small wavelengths) to  
 be almost
indistinguishable (see P. 962 in Ref. \cite{mtw}) from their analogue 
electromagnetic waves (EMW).   We, therefore theoretically,  show, using the 
known methods of 
optical  holography   and taking into account the very peculiar nature of 
GW,   
that it is also possible to  
reconstruct  subject gravitational waves. } 

     \end{abstract}

 \noindent    \underline{Pacs Numbers}: 42.40.-i, 42.40.Eq, 04.30.-w, 04.30.Nk 
     \bigskip    
     
     \markright{INTRODUCTION}
\protect \section{\bf INTRODUCTION}
The theory of electromagnetic (optical light, $X$-rays, $\gamma$-rays) 
and matter (electrons, atoms)  holography is well
established (see, for example,
\cite{Gabor,Collier,Tegze,Korecki,Barton,Han,Harp}). Theoretical and 
experimental set-ups have
become possible not only for the holographic reconstruction of large macroscopic
objects in the optical domain \cite{Gabor,Collier,Kogelnik} but also for the
microscopic resolution and imaging of minute objects such as molecules and 
atoms. What makes this imaging possible is the advancement of the early 
 holography \cite{Gabor,Collier,Kogelnik} first
to the $X$-ray \cite{Tegze} and $\gamma$-ray \cite{Korecki} domains and then to
the generation and application of holograms by using matter waves such as
electron emission from atoms
\cite{Barton,Han,Harp,Szoke,Spence,Soroko}.  \par
  In this work we wish  to further, theoretically,  
 expand and enlarge the diversity of waves  used to record and reconstruct an
 initial subject. We use in this context gravitational waves which are so unique
 and different compared to  electromagnetic and 
 matter waves. First their  interaction with
 matter is so weak that no direct \cite{Note} experimental set-up, even the giant  LIGO
 \cite{Ligo}, VIRGO \cite{Virgo}, GEO \cite{Geo} and TAMA \cite{Tama} 
  facilities have succeeded
 up to now to directly detect them (see, for example a null result report
 \cite{Abbott} of a mutual search of LIGO and GEO for GW from the pulsar
 J1939+2134).  
Second,  these waves, which are "ripples of curvature" \cite{mtw}, 
 influence the space-time  through which they proceed by further curving it 
so as to increase or decrease the  interval between the geodesics travelled by
test particles \cite{mtw}. 
Thus, holography which is thought to result from the diffraction and changes 
 of the form of the passing waves by  solid spatial objects may be discussed
 also from the point of view as if these diffraction and form changes  result from
 passing through a region of spacetime in which the curvature is stronger than
 other regions (see Figure 2). \par 
 The former discussion leads to the realization that one may diffracts and
 changes the form of an  EMW passing through 
  some finite region of space-time by two equivalent ways; 
 (1)  by some solid object placed in this region and (2)  by a strong 
   curvature (stronger than its values in neighbouring regions) imprinted in
   this region by GW. Moreover, one may,  theoretically, obtain very similar
  diffractions for these two cases if he can adjust the external form of the 
  spatial object to be such that an EMW which encounters it  will be changed 
   in the same manner as if it have  passed  through the region of strong 
   curvature. 
  Note that the principle  of finite region with stronger
 curvature than other regions stands at the basis of all the different efforts 
  made for  detecting 
 GW  from   the early mechanical bars of Weber \cite{Weber} to  the later
 mentioned interferometric detectors \cite{mtw}. \par 
 Thus,  one may discuss holography, as  done here, by considering 
 regions of stronger curvature (compared to neighbouring regions) 
 without having to place in these regions any solid spatial object. 
 We note in this context
 that already in the early holography \cite{Collier} the presence of a realistic
 solid objects were thought in certain cases  to be unnecessary for recording 
real  holograms such 
 as, for example,  in the computer-generated holograms \cite{Collier,Brown}.
 \par 
 We must note that we neither try here to find the most general and complete
 theory of possible  GW  holography  nor to discuss the fundamental aspects of 
 the gravitational 
 field as, for example, done in the works of Finkelstein {\it et al} 
 (see, for example \cite{Finkelstein}). 
 We use the remarked property emphasized in \cite{mtw} that the GW, 
  under certain conditions, is
 indistinguishable from the EMW to also discuss, at least
 theoretically, a possible GW holography. For this it is sufficient 
  to discuss plane GW's in the
 simplified transverse-traceless (TT) gauge \cite{mtw} where these waves are
 purely  spatial \cite{mtw}. \par 
 Thus, following the conventional holography \cite{Collier},  
 which necessitates a second reference
 wave which do not touch the solid object, we assume here another GW, denoted 
 $R$, which do not pass through the region passed by the
 former GW (called $S$ for subject) and serves as a reference to $S$. The two waves
 $S$ and $R$ are supposed to meet and interfere in a second space-time region
 which serves as a hologram just as the subject and reference waves in optical
 holography meet and interfere in the hologram. \par Also, as in  
 optical hologram
 \cite{Collier} one may assume that 
  the gravitational hologram is formed by the exposure (interference
  \cite{Jenkins} of $S$
 and $R$) and the duration of it. But in 
 contrast to the holograms recorded by EMW  which are solid spatial objects 
 made by altering, through exposure,  the   transmission or 
 absorption properties of the recorded
 materials  \cite{Collier} (and include the plane and volume  
 photosensitive and photographic emulsion holograms \cite{Collier}) here the
 hologram can not be a similar solid object which records the interference
 of  $S$ and $R$. This is because, as mentioned, the effect of any GW
 is to increase space-time curvature  \cite{mtw} and this is, naturally,  
  imprinted and recorded in the space-time itself  (and not in any
 solid 3-dimensional object in it) so that a  wave (EMW) passing through this
 region is  difracted.  Thus, the related hologram is, actually, a
 finite space-time region which records the interference between the GW's $S$ and
 $R$  so that if, as in the usual holography,  the
 reference wave $R$ is later sent again through this region, as illuminator, 
   one reconstructs the
space-time changes made by the subject wave $S$ in its original region. We,
theoretically, show  that this is, actually, the case. \par  
We note in 
this context that, in
 contrast to optical holography which records and reconstructs a 3-dimensional
 solid object where the  temporal evolution 
is generally
everaged and neglected \cite{Collier} the case for the later microscopic
holography is different. This is so, especially, for matter waves such as
electrons which must be discussed in quantum terms \cite{Ayman} for which time evolution is
very important \cite{Schiff}. It has been shown, for example, in \cite{Soroko}
that the discussion of holograms made by matter waves has effects similar to
those resulting from volume holograms \cite{Collier,Kogelnik} except for
replacing the spatial third dimension with the time variable.  This emphasis of
the time evolution is, especially, valid for the  holography
discussed here where, as described,  the passing GW  acts directly on the
space-time medium itself and not on any spatial object in it.  We, therefore, 
 emphasize these temporal changes and assume
that the spatial components of the finite space-time region in which the waves
$S$ and $R$ meet and interfere is very small (the small thin area $A$ 
in Figure 1).\par
In Section II we use the linearized weak field theory and introduce the relevant
subject and reference GW together with their appropriate polarization
components. In Section III we calculate the relevant intensities and the
exposure. In Section IV we represent the hologram transmittance over the small
area $A$ and calculate the required reconstructed wave which will be found to be
   proportional to the original subject wave $S$.  We conclude and summarize the
obtained results in a Concluding Remarks Section. Also, since GW  are, 
as mentioned, 
 indistinguishable, under certain
conditions, from EMW  we use  some known optical  
coherence   expressions \cite{Born} which are  introduced in a 
separate Appendix.

\markright{THE SUBJECT AND REFERENCE GW.....}

\protect \section{\bf The subject and reference GW and their polarizations }

Figure 1 shows a schematic representation of the arrangement  used in this
 discussion. In this set-up array  we assume an initial GW which have been
 detected at the point $C$ maybe by one or some collaboration of the mentioned
 interferometric detectors. This wave may be assumed 
 to be divided, through a future technology,  into two components which propagate to the two different regions 
 denoted in Figure 1 as $S$ and $R$. From these two regions the relevant GW's,
 denoted also as $S$ and $R$, propagate to the small region $A$ where they meet
 and interfere.    
As mentioned, we use the linearized weak field approximation \cite{mtw,Bergmann} 
of general relativity which although refers to the surrounding space-time 
as if it were flat (as in 
special relativity) it, nevertheless, discuss experiments and their evolutions
in a curved space-time formalism. In this theory the metric tensor components
are given by \cite{mtw} 

\begin{equation} g_{\mu\nu}=\eta_{\mu\nu}+h_{\mu\nu}+O([h_{\mu\nu}]^2),
\label{e1} \end{equation} 
where $\eta_{\mu\nu}$ is the Lorentz metric of special relativity 
\cite{mtw,Bergmann} and $h_{\mu\nu}$ is a small perturbation. This 
 $h_{\mu\nu}$  is identified with GW \cite{mtw,Thorne}
 which is itself a propagating perturbation of space-time \cite{mtw,Thorne}. 
 As known
\cite{mtw}, one of the simplest gauges to which one may subject the tensor 
$h_{\mu\nu}$ is the transverse-traceless gauge (TT) in which $h_{\mu\nu}$ has
the smallest number of components \cite{mtw,Thorne}. This is because in this gauge
\cite{mtw,Thorne}
 $h_{\mu\nu}$ is purely spatial so $h_{0\mu}=0$ and   it is also 
 transverse to the
 direction of its propagation so $h_{ij,j}=0$. Its tracelessness introduces the
 additional condition of $h_{jj}=0$.  Thus, 
 the gravitational wave is traditionally signified \cite{mtw} as 
 $h^{TT}_{\mu\nu}$ which
 is the tensor $h_{\mu\nu}$ in the $TT$ gauge.   We take into account that 
 $h^{TT}_{\mu\nu}$ is, as mentioned, purely spatial so we  
 follow   the 
 traditional holographic notation and denote the relevant subject and reference
 waves by $S(x,y,z,t)$ and $R(x,y,z,t)$ respectively. \par We assume that   
  $S(x,y,z,t)$ and $R(x,y,z,t)$  are plane waves propagating along the 
  respective vectors 
   ${\bf n}_s$ and ${\bf n}_r$ and denote the two orthogonal directions which
   are perpendicular to ${\bf n}_s$ by ${\bf e}_{s_1}$ and ${\bf e}_{s_2}$ and 
   those  perpendicular to
   ${\bf n}_r$ by ${\bf e}_{r_1}$ and ${\bf e}_{r_2}$. Thus, following the
   notation in \cite{mtw} (where the discussion there refers  to  propagation
   along the $z$ axis (see Chapter 35 there)) we denote the two unit linear 
   polarization tensors of
    $S(x,y,z,t)$ as ${\bf e}_{+_s}$,  ${\bf e}_{\times_s}$ and those of 
   $R(x,y,z,t)$ as ${\bf e}_{+_r}$,  ${\bf e}_{\times_r}$ and write 
   \begin{eqnarray} && {\bf e}_{+_s} ={\bf e}_{s_1}\otimes 
   {\bf e}_{s_1}-{\bf e}_{s_2}\otimes {\bf e}_{s_2}, \ \ \ \ 
  {\bf e}_{x_s}= {\bf e}_{s_1}\otimes 
   {\bf e}_{s_2}+{\bf e}_{s_2}\otimes {\bf e}_{s_1} \label{e2} \\ && 
   {\bf e}_{+_r} ={\bf e}_{r_1}\otimes 
   {\bf e}_{r_1}-{\bf e}_{r_2}\otimes {\bf e}_{r_2}, \ \ \ \ 
  {\bf e}_{x_r} ={\bf e}_{r_1}\otimes 
   {\bf e}_{r_2}+{\bf e}_{r_2}\otimes {\bf e}_{r_1} \nonumber
   \end{eqnarray} 
   where $\otimes$ is the tensor product. 
   Note that each GW have, like its EMW analogue,  two polarizations.  Thus, if,
   for example, the propagating GW advances vertically through an
   interferometric detector 
   then one polarization, actually, describes the known tidal forces \cite{mtw} 
   which
   oscillate along the directions \cite{Thorne} of east-west and north-south.  
   The other
   polarization describes those tidal forces which oscillate along the
   directions \cite{Thorne} of northeast-southwest and northwest-southeast. 
   In the following we assume  
  the subject and reference waves    to be  
  polychromatic so their  sources  emit light at 
  several frequencies. 
  We denote by  ${\bf r}$  the position  vector  of a point
  in space and  signify   the respective 
  direction  
cosines of ${\bf n}_s$,  ${\bf n}_r$  by $\cos(\alpha_s), \cos(\beta_s), \cos(\gamma_s)$ 
and $\cos(\alpha_r), \cos(\beta_r), \cos(\gamma_r)$.  Thus, taking into account 
that $k=\frac{2\pi}{\lambda}$ and defining the spatial frequencies
 $\xi_s=\frac{\cos(\alpha_s)}{\lambda}$, 
  $\eta_s=\frac{\cos(\beta_s)}{\lambda}$, 
  $\zeta_s=\frac{\cos(\gamma_s)}{\lambda}$,  
 $\xi_r=\frac{\cos(\alpha_r)}{\lambda}$, 
  $\eta_r=\frac{\cos(\beta_r)}{\lambda}$, 
  $\zeta_r=\frac{\cos(\gamma_r)}{\lambda}$ 
   one
  may write, for example,   the subject and reference  GW  as  
\begin{eqnarray} && S(x,y,z,t)=\Re[ (A_{+_s}{\bf e}_{+_s}+A_{\times_s}{\bf e}_{\times_s})
e^{ik{\bf r}\cdot {\bf n}_s}(c_0e^{i2\pi ft}+
c_1e^{i2\pi(f+\epsilon_1)t} + \nonumber \\ && + c_2e^{i2\pi(f+\epsilon_2)t}
+\cdots)\big]=
\Re[ (A_{+_s}{\bf e}_{+_s}+A_{\times_s}{\bf e}_{\times_s})
\exp[ik(x\cos(\alpha_s)+ \label{e3} \\ && 
+y\cos(\beta_s)+z\cos(\gamma_s)]
e^{i2\pi ft}\sum_ic_ie^{i2\pi\epsilon_it}]=
\Re[ (A_{+_s}{\bf e}_{+_s}+A_{\times_s}{\bf e}_{\times_s}) \cdot \nonumber \\ && 
\cdot \exp[i2\pi(\xi_s x+ \eta_s y+ \zeta_s z)]
e^{i2\pi ft}\cdot 
{\bf g}(t)] \nonumber   
\end{eqnarray} 
\begin{eqnarray}  &&  R(x,y,z,(t+\tau))=  \Re[
(A_{+_r}{\bf e}_{+_r}+A_{\times_r}{\bf e}_{\times_r})e^{ik{\bf r}\cdot {\bf n}_r}
(c_0e^{i2\pi f(t+\tau)}+
c_1e^{i2\pi(f+\epsilon_1)(t+\tau)}+ \nonumber \\ && + 
c_2e^{i2\pi(f+\epsilon_2)(t+\tau)}+\cdots)] = 
 \Re[ (A_{+_r}{\bf e}_{+_r}+A_{\times_r}{\bf e}_{\times_r})
 \exp[ik(x\cos(\alpha_r)+y\cos(\beta_r)+ \label{e4} \\ && + z\cos(\gamma_r)]\cdot 
  e^{i2\pi
 f(t+\tau)}\sum_ic_ie^{i2\pi\epsilon_i(t+\tau)}] =
\Re[ (A_{+_r}{\bf e}_{+_r}+A_{\times_r}{\bf e}_{\times_r})
\cdot \exp[i2\pi(\xi_r x+ \nonumber \\ && + \eta_r y +\zeta_r z)]\cdot 
e^{i2\pi f(t+\tau)}\cdot 
{\bf g}(t+\tau)] \nonumber 
  \end{eqnarray}
where $\Re$ denotes the real part of the following complex expression. 
  The amplitudes $A_{+_s}$, $A_{\times_s}$ and 
$A_{+_r}$, $A_{\times_r}$ refer respectively to the 
 modes of polarizations ${\bf e}_{+_s}$, ${\bf e}_{\times_s}$ and 
$e_{+_r}$, $e_{\times_r}$. In the following, for ease of notation, we denote 
$s_0=A_{+_s}{\bf e}_{+_s}+A_{\times_s}{\bf e}_{\times_s}$ and $r_0=A_{+_r}{\bf 
e}_{+_r}+
A_{\times_r}{\bf e}_{\times_s}$. The parameter $\tau$ in Eq (\ref{e4}) 
is defined 
by $c\tau$, where  
$c$ is the velocity of the GW which is equal to the velocity
of light, so  that $c\tau$ is the  path difference between the
paths of  $S$ and  $R$  as they propagate from their 
places at $S$ and $R$ (see Figure 1)
to the small area $A$. At the last results of Eqs (\ref{e3})-(\ref{e4}) we have
used the definitions  ${\bf g}(t)=\sum_ic_ie^{i2\pi\epsilon_it}$ and 
${\bf g}(t+\tau)=\sum_ic_ie^{i2\pi\epsilon_i(t+\tau)}$ where we assume that
since $S$ and $R$ have common source (represented by the point $C$ in Figure 1)
the quantities $\epsilon_i$ and the coefficients $c_i$ are the same in 
${\bf g}(t)$ and ${\bf g}(t+\tau)$. \par 
 The subject wave $S$
from Eq (\ref{e3}) may be decomposed into a component, denoted ${\bf S}_{=}$, 
 which is polarized 
parallel to the polarization  direction of the reference wave $R$ and another
components, denoted $ S_{+}$, which is perpendicular to this direction.
Thus, denoting the angle between the polarization directions of $S$ 
and $R$ from
Eqs (\ref{e3})-(\ref{e4}) 
 by  $W$ (which is the same as that between the propagating rays $S$ and $R$
 (see Figure 1 and the text after Eqs (\ref{e5})-(\ref{e6}))  one
may write $S_{=}$ and $S_{+}$ as 
\begin{equation} S_{=}=\Re[ s_0\cdot 
\exp[i(2\pi \xi_s x+2\pi \eta_s y+2\pi \zeta_s z)]e^{i2\pi ft}\cdot 
{\bf g}(t)\cdot \cos(W)] \label{e5} \end{equation}
\begin{equation} S_{+}=\Re[ s_0\cdot 
\exp[i(2\pi \xi_s x+2\pi \eta_s y+2\pi \zeta_s z)]e^{i2\pi ft}\cdot 
{\bf g}(t)\cdot \sin(W)] \label{e6} \end{equation}

\begin{figure}
\centerline{
\epsfxsize=5.5 in
\begin{turn}{-90}
\epsffile{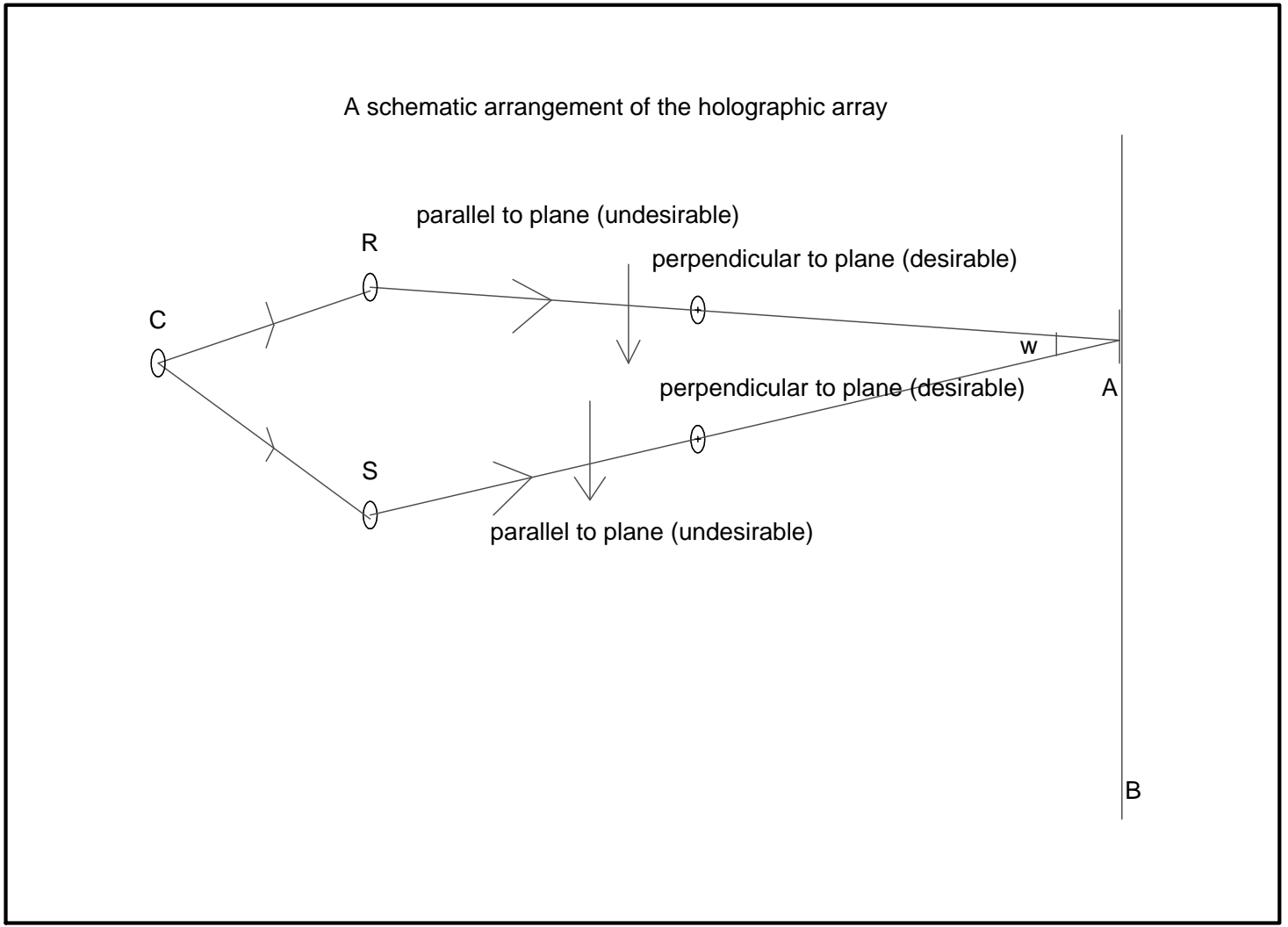}
\end{turn}}
     \caption{The subject and reference waves are shown as rays originating 
      at their
     common source at $C$  from there they advance first 
     to their respective  points $S$ and $R$ 
     and then  to the small area $A$. It is also shown, for each wave, 
      the two perpendicular components which are parallel and perpendicular to
      the plane of the figure and which denote the directions of polarization.
      The undesirable and desirable components of these polarizations (see text) 
      are also shown.    } 
      \end{figure}

In Figure 1 we show not only the propagating GW's of $S$ and $R$ but also the
two components, for each GW, which are parallel and perpendicular to the plane
of the figure. These components serve as polarization vectors. Note, however,
that the angle between the polarization components of $S$ and $R$ which are
parallel to the plane of Figure 1 is $W$ (which equals the angle between the
propagating  $S$ and $R$ (see Figure 1)) whereas the angle between the
polarization components perpendicular to this plane is zero. Thus,  refering to
the later components one may realize from Eqs (\ref{e5})-(\ref{e6}) that the
component $S_{+}$  is zero whereas $S_{=}$  is maximum. In other words, 
 the desirable
components of polarization are those perpendicular to Figure 1 as written
explicitly in this figure. Note that this criterion applies also for optical
holography \cite{Collier}. We continue to use in the text the angle $W$ since we
are, especially, interested in the intensities of the GW and for this, as
realized from the following section,  one may
obtain the same result regardless if he uses the perpendicular or the parallel
components of polarization. \par
The effect of the polarizing tensors of either the gravitational plane wave 
 $S$ or $R$  upon the 
space-time medium
is best understood from Figure 2 which, actually, shows the left half of Figure
35.2 in \cite{mtw}. This figure  shows how a closed circular (elliptic)
array of test particles are changed, due to the resulting increased curvature, 
 to
elliptic (circular) array. These changes, as seen from the figure,  are periodic
and their exact form depend upon the  values of the phase (shown in degrees 
 at the right hand side 
the figure) and upon  the unit linear polarization tensor.

\begin{figure}
\centerline{
\epsfxsize=5.5 in
\begin{turn}{-90}
\epsffile{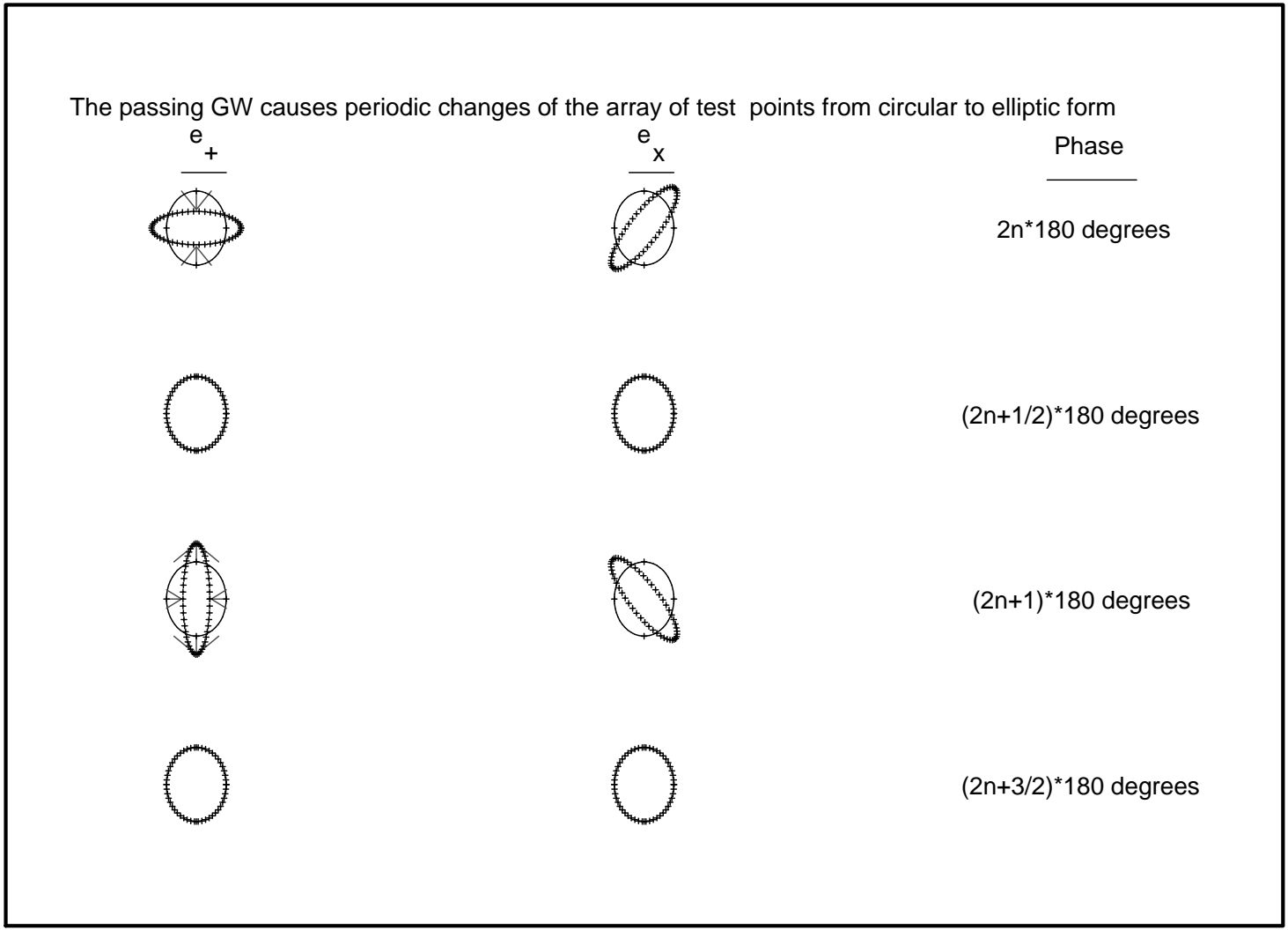}
\end{turn}}
     
     \caption{The figure shows the effect of a GW passing through a region in
     which  some test particles are shown 
     arrayed in a closed form. The presence of the GW  
     causes the space-time in this region  to be more curved than usually 
     when it is
     absent  and this in turn  changes the closed form of the array of test
     particles from a circular (elliptic) to an elliptic (circular)  form. 
     This behaviour, for the gravitational plane wave discussed here, 
      is repeated in a periodic  fashion and  depends  upon the values 
      assumed by
      the phase as shown at the right (in degrees) and 
      upon the corresponding nature of the
     polarization tensor $e_x$ or $e_+$.   }
     \end{figure}
   
\bigskip 
\markright{THE INTENSITIES AND EXPOSURE.....}

\protect \section{The intensities and exposure of the GW}

The separate intensities of the  waves $R(x,y,(t+\tau))$, $S_{=}$  and $S_{+}$, 
denoted  $I_{R}(t+\tau)$, $I_{S_=}(t)$ and $I_{S_+}(t)$, 
at the small area $A$ are found from Eqs (\ref{e4}), (\ref{e5})-(\ref{e6}) 
as follows 
\begin{equation} I_{R}(t+\tau)=<\!R(x,y,(t+\tau))R^*(x,y,(t+\tau))\!>=
r^2_0 <\!{\bf g}(t+\tau){\bf g^*}(t+\tau)\!> \label{e7} \end{equation} 
\begin{equation} I_{S_=}(t)=<\!S_=(t)S^*_=(t)\!>=
s^2_0 <\!{\bf g}(t){\bf g^*}(t)\!>\cos^2(W) \label{e8} \end{equation} 
 \begin{equation} I_{S_+}(t)=<\!S_+(t)S^*_=(t)\!>=
s^2_0 <\!{\bf g}(t){\bf g^*}(t)\!>\sin^2(W) \label{e9} \end{equation} 
The overal intensity at the area $A$ of the waves  
 $R(x,y,(t+\tau))$, $S_{=}$, and $S_{+}$ is the sum of the separate 
 intensities from
Eqs (\ref{e7})-(\ref{e9}) plus the interference formed by these waves. But we
must note that no interference is formed from  $S_+$  and the polarization
vector of $R(x,y,(t+\tau))$  because, as mentioned, they are  perpendicular 
to each other. 
Thus, for interference we should take only the interaction of $S_=$ and the
polarization direction of 
$R(x,y,(t+\tau))$  
 which are  parallel to each other. In other words, the
required total intensity at the small area $A$ is
\begin{eqnarray} && I_{total}=I_{R}(t+\tau)+I_{S_=}(t)+I_{S_+}(t)+
2\Re [<\!R(x,y,(t+\tau))S^*_{=}\!>]=
\nonumber \\ && = r^2_0 <\!{\bf g}(t+\tau){\bf g^*}(t+\tau)\!>+
s^2_0 <\!{\bf g}(t){\bf g^*}(t)\!>\cos^2(W)+ s^2_0 <\!{\bf g}(t){\bf g^*}(t)\!>
\cdot \label{e10} \\ && \cdot \sin^2(W)+  
2\Re [r_0s_0\cos(W)\cdot \exp[i2\pi((\xi_r -\xi_s)x+(\eta_r-\eta_s)y+(\zeta_r-\zeta_s)z)]
\cdot \nonumber \\ && \cdot
e^{i2\pi f\tau}<\!{\bf g}(t+\tau){\bf g^*}(t)\!>]  \nonumber 
\end{eqnarray}
From Eq $(A7)$ in the Appendix  one may realize \cite{Collier}
that since $|{\bf {\hat \mu}}_T(\tau)|=|{\bf \mu}_T(\tau)|$ where 
${\bf {\hat \mu}}_T(\tau)= {\bf \mu}_T(\tau)e^{-i2\pi f\tau}= \frac{<\!{\bf g}(t+\tau){\bf g}^*(t)\!>}
{<\!{\bf g}(t){\bf g}^*(t)\!>} $ it is possible to 
 write 
${\bf {\hat \mu}}_T(\tau)=|{\bf \mu}_T(\tau)|\cdot e^{i\zeta(\tau)}$
where  $e^{i\zeta(\tau)}$ is a phase factor. Thus,  using  the last equation  
 one may write 
the total intensity
from Eq (\ref{e10})  as \cite{Collier} 
\begin{eqnarray} && I_{total}=r^2_0 <\!{\bf g}(t+\tau){\bf g^*}(t+\tau)\!>+
s^2_0 <\!{\bf g}(t){\bf g^*}(t)\!>+  
2\Re[r_0s_0\cos(W)\cdot \label{e11} \\ && \cdot 
\exp[i2\pi((\xi_r -\xi_s)x+(\eta_r-\eta_s)y+(\zeta_r-\zeta_s)z)]
e^{i2\pi f\tau}
|{\bf \mu}_T(\tau)|\cdot   e^{i\zeta(\tau)}\cdot \nonumber \\ && \cdot 
<\!{\bf g}(t){\bf g^*}(t)\!>] =r^2_0<\!{\bf g}(t+\tau){\bf
g^*}(t+\tau)\!>+s^2_0<\!{\bf g}(t){\bf
g^*}(t)\!>+ \nonumber \\ && + 2r_0s_0\cos(W) \cdot |{\bf \mu}_T(\tau)|
\cos(\beta(x,y,z,\tau))  <\!{\bf g}(t){\bf
g^*}(t)\!>, \nonumber \end{eqnarray} 
 where $\beta(x,y,z,t)=2\pi[(\xi_r -
 \xi_s)x+(\eta_r-\eta_s)y+(\zeta_r-\zeta_s)z+f\tau]+\zeta(\tau)$.
 The intensity from Eq (\ref{e11}) is recorded on the hologram  through
 exposure $E$ which is assumed to be, in analogy with optical holograms, 
 proportional  \cite{Collier} to  the product of the 
 intensity $I_{total}$ and the
 exposure time $\tau_e$. That is, denoting the proportionality constant by $C$ 
   one may write, using  Eq (\ref{e11}),  
 the exposure as 
 \begin{eqnarray}  && E(x,y,z,t)=C\tau_eI_{total}=
 C\tau_e [s^2_0<\!{\bf g}(t){\bf
g^*}(t)\!>+r^2_0<\!{\bf g}(t+\tau){\bf
g^*}(t+\tau)\!> \label{e12} \\ && +  
2r_0s_0\cos(W)|{\bf \mu}_T(\tau)|<\!{\bf g}(t){\bf
g^*}(t)\!>
\cos(\beta(x,y,z,\tau))]=E_0+E_1(x,y,z,t), \nonumber  \end{eqnarray} 
where $$E_0= C\tau_e(s^2_0<\!{\bf g}(t){\bf
g^*}(t)\!>+r^2_0<\!{\bf g}(t+\tau){\bf
g^*}(t+\tau)\!>)$$   $$E_1(x,y,z,t)=
2C\tau_es_0r_0\cos(W)|{\bf \mu}_T(\tau)| 
<\!{\bf g}(t){\bf g^*}(t)\!>
\cos(\beta(x,y,z,\tau))$$ 

\markright{THE HOLOGRAM TRANSMITTANCE AND......}

\protect \section{The hologram transmittance and the reconstructed GW}

Referring  to the former equations one may realize that if the expression  
$\frac{r^2_0}{s^2_0}$   satisfies 
$\frac{r^2_0}{s^2_0}>1$  over the small area
$A$  then one also have $\frac{r^2_0}{s^2_0}>\frac{r_0}{s_0}$ 
and consequently $E_0>E_1$ over this area $A$ of the hologram. In such 
case, analogously to optical holography \cite{Collier},  
one may write the hologram transmittance over the area $A$ as a Taylor series
\cite{Collier}
\begin{equation} {\bf t}_E={\bf t}_E(E_0)+E_1\frac{d{\bf
t}_E}{dE}|_{E_0}+\frac{1}{2}E_1^2\frac{d^2{\bf t}_E}{dE^2}|_{E_0}+\cdots 
\label{e13} \end{equation}  
 In optical holography this representation of the  
transmittance  is general and includes the possibility of either 
amplitude or phase modulation by the
hologram \cite{Collier}.
  Now, (1): we 
assume that all the coefficients of second and higher order terms in 
Eq (\ref{e13}) 
$\frac{d^2{\bf t}_E}{dE^2}|_{E_0}, \frac{d^3{\bf t}_E}{dE^3}|_{E_0}, ...$  
are negligible and (2): that the factor 
$\cos(\beta(x,y,z,\tau))$ from Eq (\ref{e12}) is written as a sum of 
 exponentials from which the term $\frac{1}{2}e^{-i\beta(x,y,z,\tau)}$ is 
 chosen (as done in optical holography \cite{Collier})  
 where $\beta$ is given by the inline equation after Eq (\ref{e11}).
 Thus,  for
 reconstructing the subject wave $S(x,y,z,t)$ from Eq (\ref{e3}) one illuminates
the hologram ${\bf t}_E$ with the reference wave $R(x,y,z,(t+\tau))$ from Eq
(\ref{e4}) so that the GW obtained  from the exposure $E_1$ (the
constant $E_0$ has no role in this reconstruction) 
at  the small area $A$ is 
\begin{eqnarray} && W_r(x,y,z,t)=R(x,y,z,(t+\tau)){\bf t}_E=r_0\cdot 
\exp[i(2\pi(\xi_r x+2\pi \eta_r y+2\pi \zeta_r z))]\cdot \nonumber \\ && \cdot
e^{i2\pi f(t+\tau)} {\bf g}(t+\tau)E_1\frac{d{\bf
t}_E}{dE}|_{E_0}=C\tau_er_0^2s_0\cos(W)|{\bf \mu}_T(\tau)|
<\!{\bf g}(t){\bf g^*}(t)\!> \cdot \nonumber \\&& \cdot
e^{-i(\beta(x,y,z,\tau))}\cdot 
\exp[i(2\pi(\xi_r x+2\pi \eta_r y+2\pi \zeta_r z))] \cdot
e^{i2\pi f(t+\tau)} {\bf g}(t+\tau)\frac{d{\bf
t}_E}{dE}|_{E_0}= \label{e14} \\ && = C\tau_er_0^2s_0\cos(W)|{\bf \mu}_T(\tau)|
<\!{\bf g}(t){\bf g^*}(t)\!>\exp[i2\pi(\xi_sx +\eta_sy+\zeta_sz)]e^{-i\zeta(\tau)}
\cdot \nonumber \\ && \cdot e^{i2\pi ft}{\bf g}(t+\tau)\frac{d{\bf
t}_E}{dE}|_{E_0}
\nonumber \end{eqnarray}
In order to continue in our analytical reconstruction of the subject wave
$S(x,y,z,t)$ we first  show that $ {\bf g}(t+\tau)\cdot <\!{\bf g}(t){\bf g}^*(t)\!>=
{\bf g}(t)\cdot <\!{\bf g}(t+\tau){\bf g}^*(t)\!>$. 
 This is done by   taking into account that ${\bf
 g}(t)=\sum_ic_ie^{i2\pi\epsilon_it}$, \ \ \  
 ${\bf g}(t+\tau)=\sum_ic_ie^{i2\pi\epsilon_i(t+\tau)}$ (see the
 discussion after Eq (\ref{e4}))  and 
 $<\!{\bf g}(t){\bf g}^*(t)\!>=\lim_{T \to \infty}\frac{1}{2T}
\int_{-T}^{T}{\bf g}(t){\bf g}^*(t)dt$, \ \ \ $<\!{\bf g}(t+\tau){\bf g}^*(t)\!>=
\lim_{T \to \infty}\frac{1}{2T}
\int_{-T}^{T}{\bf g}(t+\tau){\bf g}^*(t)dt$ (see Eq $(A5)$ in the 
Appendix). Thus, integrating the elementary exponentials and taking the
corresponding limits $T \to \infty$ one obtains the following results
\begin{equation} <\!{\bf g}(t){\bf g}^*(t)\!>=\sum_i|c_i|^2, \ \ \   
<\!{\bf g}(t+\tau){\bf g}^*(t)\!>=\sum_i|c_i|^2e^{i2\pi\epsilon_i\tau} 
\label{e15} \end{equation}
From the last equations, the definitions of ${\bf g}(t)$, \ 
${\bf g}(t+\tau)$  and  the discussion after Eq (\ref{e4}) 
about  $\epsilon_i$ and  $c_i$ which 
are the same in 
${\bf g}(t)$ and ${\bf g}(t+\tau)$ one may realize that 
\begin{eqnarray} && {\bf g}(t+\tau)\cdot <\!{\bf g}(t){\bf g}^*(t)\!>=
\sum_i|c_i|^2e^{i2\pi\epsilon_i(t+\tau)}\cdot \sum_i|c_i|^2=
\label{e16} \\ && =
\sum_i|c_i|^2e^{i2\pi\epsilon_it}\cdot \sum_ie^{i2\pi\epsilon_i\tau}|c_i|^2=
{\bf g}(t)\cdot <\!{\bf g}(t+\tau){\bf g}^*(t)\!>, \nonumber \end{eqnarray}
which is what we set to prove.  Using the last equation and Eq $(A7)$   
in the Appendix (from which we see, as mentioned after Eq (\ref{e10}),   
that the equality  $|{\bf {\hat \mu}}_T(\tau)|=|{\bf \mu}_T(\tau)|$ 
leads to  
${\bf {\hat \mu}}_T(\tau)=|{\bf \mu}_T(\tau)|\cdot e^{i\zeta(\tau)}$)  
one may write   the
reconstructed wave from Eq (\ref{e14}) as 
\begin{eqnarray}  && W_r(x,y,z,t)=C\tau_er^2_0s_0 
\cos(W)|{\bf \mu}_T(\tau)|
<\!{\bf g}(t+\tau){\bf g}^*(t)\!> 
\exp[i2\pi(\xi_sx + \nonumber \\ && +\eta_sy+\zeta_sz)]  
\cdot e^{i2\pi ft}e^{-i\zeta(\tau)}
\cdot {\bf g}(t)\frac{d{\bf
t}_E}{dE}|_{E_0}=C\tau_er^2_0s_0 
\cos(W)|{\bf \mu}_T(\tau)|^2 \cdot \label{e17} \\ && \cdot 
<\!{\bf g}(t){\bf g}^*(t)\!>e^{i2\pi ft} \cdot 
\exp[i2\pi(\xi_sx +\eta_sy+\zeta_sz)]   
\cdot {\bf g}(t)\frac{d{\bf
t}_E}{dE}|_{E_0}
\nonumber  \end{eqnarray}
Now, taking into account our neglection (see the discussion after Eq  
(\ref{e13})) of the coefficients of the 
higher order terms in Eq (\ref{e13}) 
$\frac{d^2{\bf t}_E}{dE^2}|_{E_0}, \frac{d^3{\bf t}_E}{dE^3}|_{E_0}, ...$ 
 one may realize that if 
$\frac{d^2{\bf t}_E}{dE^2}|_{E_0}=0$  then  $\frac{d{\bf
t}_E}{dE}|_{E_0}=constant$. Thus, using the last result, the definition of $s_0$
as given after Eq (\ref{e4}), and the first of Eqs 
(\ref{e15}) one may write the reconstructed wave from Eq (\ref{e17}) as 
\begin{eqnarray}  && W_r(x,y,z,t) = constant\cdot s_0\cdot {\bf g}(t)e^{i2\pi ft}
\exp[i2\pi(\xi_sx +\eta_sy+\zeta_sz)] = \label{e18} \\ && =constant 
\cdot (A_{+_s}{\bf e}_{+_s}+A_{\times_s}{\bf e}_{\times_s})
{\bf g}(t)e^{i2\pi ft}
\exp[i2\pi(\xi_sx +\eta_sy+\zeta_sz)]= \nonumber \\ && =
 constant\cdot S(x,y,z,t), \nonumber
\end{eqnarray}
where $S(x,y,z,t)$ is the subject wave given by Eq (\ref{e3}). Thus, as in
optical holography, we see 
that the original subject wave has been reconstructed.

     \markright{CONCLUDING REMARKS}
     
\protect  \section{Concluding Remarks}
We have discussed  gravitational wave holography in which all
the involved waves; subject, reference and illumnator are gravitational waves. 
The  nature of these waves, compared to their electromagetic analogues,
causes the resulting holography to be somewhat unique. First,  the
interaction of these waves with matter is so weak that no experimental set-up
have, up to now, succeeded to directly \cite{Note} detect them. Second, these
waves act upon the space-time structure itself by increasing its curvature so
that 
any wave (for example, EMW) which passes in this region  undergoes similar changes as those 
occuring when encountering 
a corresponding solid spatial object. 
That is, the same diffraction and form changes in some finite region 
may result from
either a strong space-time curvature in it compared to other neighbouring
regions or from a corresponding suitably designed solid object. Note that
this is reminiscent of the famous Einstein "elevator" \cite{Bergmann}  in which
a man closed inside this accelerating cabin can not be sure if this   
accelaration  is due to 
 a gravitational force all around or maybe he is 
  in a region absent of any gravitation and  that other force  
 pulls the cabin  with the known attraction of gravity. \par   
 The strong curvature imprinted by the GW upon the
space-time medium remains in this medium \cite{mtw} even after the wave have 
completely passed away  \cite{mtw} especially if this  
GW is strong enough 
or if it has passed this region a large number of times \cite{mtw}.  
\par
In our discussion  we have used the known methods of optical
holography \cite{Gabor,Collier} and, especially, the realization \cite{mtw} 
that, under certain conditions such as very small wavelength, one can not,
theoretically, diffentiate between GW and EMW.    
We have, thus,  shown that passing a reference GW  (not in the same
region passed by the subject wave) and letting these two waves meet and
interfere in some other  space-time region (hologram) then if this reference 
wave is 
again passed, as the corresponding EMW illuminator, 
through this hologram  the result will be a reconstruction of the subject wave. 
 \par 
Although this discussion is purely theoretical one may hope that a future
advanced technology will be developed which will enable the next generation of
scientists to use and manipulate GW the same way we are able now to use EMW.

\begin{appendix}

\markright{APPENDIX}

\section{APPENDIX}

We use here the mentioned characteristic of the almost theoretical identity
(valid under certain conditions such as very short wavelengths) 
between GW and EMW and assume, as we do in the main text, that we may use the
known results \cite{Born} regarding the spatial and (or) temporal 
coherence between two
waves. We, thus,   introduce here  some  relevant 
expressions 
\cite{Collier,Born} for the 
coherence between two
complex electric waves ${\bf v}_1$ and ${\bf v}_2$ which advance from points 
$P_1$ and $P_2$ at some 
screen  to the  point $Q$ at another. This is similar to the set-up in 
Figure 1 in which the two GW's $S$ and $R$ propagate from the corresponding points
$S$ and $R$ to the small area $A$.  The time average of the interference term between  
${\bf v}_1$ and ${\bf v}_2$ is written as \cite{Born}
$$ <\!{\bf v}_1 {\bf v^*}_2+{\bf v}^*_1 {\bf v}_2\!>=
2\Re<\!{\bf v}_1 {\bf v}^*_2\!> \eqno (A1) $$
It has been shown in \cite{Born} that the time average from Eq $(A1)$ can be
expressed in terms of the complex degree of coherence $\Upsilon_{12}(\tau)$ which
relates the correlation of  ${\bf v}_1$ and ${\bf v}_2$ at $P_1$ and $P_2$ to
the interference time average at $Q$. The parameter $\tau$ denotes the time
difference in arrival from the points $P_1$, $P_2$ to $Q$.   Thus, denoting 
by ${\bf v}(t)_{P_1}$,  ${\bf v}(t)_{P_2}$ the complex fields at $P_1$, $P_2$ and by 
$2<\!{\bf v}(t)_{P_1}{\bf v}^*(t)_{P_1}\!>$, 
$2<\!{\bf v}(t)_{P_2}{\bf v}^*(t)_{P_2}\!>$ the
corresponding light intensities one may define  the complex 
coherence $\Upsilon(\tau)$  \cite{Born} as
$$ \Upsilon_{12}(\tau)=\frac{<\!{\bf v}(t+\tau)_{P_1}{\bf v}^*(t)_{P_2}\!>}
{[<\!{\bf v}(t)_{P_1}{\bf v}^*(t)_{P_1}\!><\!{\bf v}(t)_{P_2}{\bf
v}^*(t)_{P_2}\!>]^{\frac{1}{2}}}= \eqno (A2) $$
$$   =\frac{\lim_{T \to
\infty}\frac{1}{2T}\int_{-T}^T {\bf v}(t+\tau)_{P_1}{\bf v}^*(t)_{P_2}dt}
{[(\lim_{T \to
\infty}\frac{1}{2T}\int_{-T}^T {\bf v}(t)_{P_1}{\bf v}^*(t)_{P_1}dt)
(\lim_{T \to
\infty}\frac{1}{2T}\int_{-T}^T {\bf v}(t)_{P_2}{\bf
v}^*(t)_{P_2}dt)]^{\frac{1}{2}}} $$  
The two Eqs $(A1)-(A2)$ were related in \cite{Born} through
$$ 2\Re[<\!{\bf v}_1{\bf v}^*_2\!>]=2(I_1I_2)^{\frac{1}{2}}\Re
[\Upsilon_{12}(\tau)]=2(I_1I_2)^{\frac{1}{2}}|\Upsilon_{12}(\tau)|
\cos(\beta_{12}(\tau)), \eqno (A3) $$
where $I_1$, $I_2$ are the  intensities at $Q$ from $P_1$, $P_2$ and 
$|\Upsilon_{12}(\tau)|$,  $\beta_{12}(\tau)$ are the modulus and phase of 
$\Upsilon_{12}(\tau)$. Note that $\Upsilon_{12}(\tau)$ is a measure of both  the
temporal and spatial aspects of the coherence \cite{Collier,Born}. For $\tau=0$ 
one may assume the points $P_1$, $P_2$ at screen $A$ to be at the 
$(x, y)$ plane where $P_1$ is located at the origin of this plane. Thus,
denoting $\Upsilon_{12}(0)={\bf \mu}_s(x,y)$ one may write Eq $(A2)$ for
this case as 
$$ 
{\mu(x,y)}_s=\frac{\int_{-\infty}^{\infty}{\bf v}(0,0,t){\bf v}^*(x,y,t)dt}
{[\int_{-\infty}^{\infty}{\bf v}(0,0,t){\bf v}^*(0,0,t)dt
\int_{-\infty}^{\infty}{\bf v}(x,y,t){\bf v}^*(x,y,t)dt]^{\frac{1}{2}}}
\eqno (A4) $$ Note that now, in cotrast to $\Upsilon_{12}(\tau)$ which denotes
both temporal and spatial coherence,  ${\mu(x,y)}_s$ is the complex spatial coherence of
the source  in the $(x,y)$ plane. Note also that if the waves
${\bf v}_{P_1}$, ${\bf v}_{P_2}$ propagate along the same direction from a point
source then ${\bf v}(t)_{P_1}={\bf v}(t)_{P_2}$ and ${\mu(x,y)}_s=1$ 
(see Eq $(A4)$).  In such a case one may replace in Eq $(A2)$ 
\cite{Collier}
$\Upsilon_{12}(\tau)$ by ${\bf \mu}_T(\tau)$,   equate 
${\bf v}(t)_{P_1}={\bf v}(t)_{P_2}={\bf v}(t)$, 
${\bf v}(t+\tau)_{P_1}={\bf v}(t+\tau)$ and write for ${\bf \mu}_T(\tau)$ which
is the complex temporal coherence 
$$ {\bf \mu}_T(\tau)= \frac{\lim_{T \to \infty}\frac{1}{2T}
\int_{-T}^{T}{\bf v}(t+\tau){\bf v}^*(t)dt}{\lim_{T \to \infty}\frac{1}{2T}
\int_{-T}^{T}{\bf v}(t){\bf v}^*(t)dt}  \eqno (A5) $$ 
  Substituting in the last equation  ${\bf v}(t)=S(x,y,z,t)$, where $S(x,y,z,t)$ 
is given by Eq (\ref{e3}) one obtains   
$$  {\bf \mu}_T(\tau)=e^{i2\pi f\tau}\cdot \frac{\lim_{T \to \infty}\frac{1}{2T}
\int_{-T}^{T}{\bf g}(t+\tau){\bf g}^*(t)dt}{\lim_{T \to \infty}\frac{1}{2T}
\int_{-T}^{T}{\bf g}(t){\bf g}^*(t)dt} =e^{i2\pi f\tau}
\frac{<\!{\bf g}(t+\tau){\bf g}^*(t)\!>}{<\!{\bf g}(t){\bf g}^*(t)\!>} \eqno
(A6) $$  One may solve the last equation \cite{Collier}  for 
$\frac{<\!{\bf g}(t+\tau){\bf g}^*(t)\!>}{<\!{\bf g}(t){\bf g}^*(t)\!>}$ and obtains 
$$ \frac{<\!{\bf g}(t+\tau){\bf g}^*(t)\!>}
{<\!{\bf g}(t){\bf g}^*(t)\!>}={\bf \mu}_T(\tau)e^{-i2\pi f\tau}=
{\bf {\hat \mu}}_T(\tau) \eqno (A7) $$
Eq $(A5)$ may be Fourier transformed into 
$$ {\bf \mu}_T(\tau)= 
\frac{\int_{-\infty}^{\infty}{\bf V}(f){\bf V}^*(f)\cdot e^{i2\pi f\tau}df}{
\int_{-\infty}^{\infty}{\bf V}(f){\bf V}^*(f)df}, \eqno (A8) $$ 
where ${\bf V}(f)$ is the temporal Fourier transform of ${\bf v}(\tau)$
\cite{Collier,Bracewell}.
 
\end{appendix}

\markright{REFERENCES}
\bigskip \bibliographystyle{plain}

\end{document}